\documentclass[twocolumn]{aastex631}

\usepackage{amsmath}
\usepackage{graphicx}
\usepackage{float}

\shorttitle{False Positive Analysis of TOI 864.01}
\shortauthors{Escolà-Rodrigo}

\begin{document}

\title{Vetting and False Positive Analysis of TOI 864.01: \\ Evidence for a Likely Hierarchical Eclipsing Binary Masked by Dilution}

\author[0009-0007-1479-2386]{Biel Escolà-Rodrigo}
\affiliation{Independent Researcher}

\begin{abstract}
We present a detailed vetting analysis of the TESS candidate TOI 864.01, initially identified as a potential ultra-short-period ($P \approx 0.52$ d) Earth-sized planet orbiting an M-dwarf. Using 12 sectors of TESS photometry spanning a multi-year baseline, we recover a robust periodic transit-like signal. While the recovered transit depth is attenuated by detrending ($\approx 158$ ppm), the SPOC pipeline reports an undiluted depth of $\approx 640$ ppm. Stellar characterization based on Gaia DR3 astrometry yields a nominally single-star solution (RUWE = 1.18), highlighting the limitations of astrometric vetting for tight companions. 

We performed a statistical validation analysis using \texttt{TRICERATOPS}, aggregating data from all 12 available sectors. The analysis yields a False Positive Probability (FPP) of $0.25$ and a Nearby False Positive Probability (NFPP) of $< 10^{-4}$. While these metrics ostensibly classify the target as a viable planetary candidate based on Gaia-resolved sources, they fail to account for sub-pixel companions. Crucially, archival high-resolution imaging from the TESS Follow-up Observing Program (TFOP SG1) reveals a stellar companion at a separation of 0.04", unresolved by both Gaia and TESS. When this companion is considered, the signal is best interpreted as a Hierarchical Eclipsing Binary (HEB) on the companion. Bayesian model comparison yields an inconclusive result ($\Delta \ln Z \approx 0.25$), consistent with the degeneracy introduced by unresolved blending. Ground-based follow-up photometry further supports significant dilution, with a measured transit depth ($\approx 0.37$ ppt) shallower than predicted ($\approx 0.64$ ppt) and a timing offset of 6.3 minutes. Taken together, we classify TOI 864.01 as a probable False Positive and recommend its retirement from planetary candidate lists.
\end{abstract}


\section{Introduction}

The Transiting Exoplanet Survey Satellite (TESS) has revolutionized the search for nearby exoplanets \citep{Ricker2015}. However, the mission's large pixel scale ($\sim 21''$/pixel) makes candidates highly susceptible to photometric blending. A background source or a bound stellar companion within the same pixel can mimic a planetary transit signal, leading to False Positives (FPs).

Distinguishing between true Ultra-Short Period (USP) planets and diluted Eclipsing Binaries (EBs) requires a combination of precise photometry, centroid analysis, high-resolution imaging, and statistical validation. In this work, we analyze TOI 864.01 (TIC 231728511), a candidate that illustrates the limitations of photometric validation alone when high-contrast neighbors are present. This case demonstrates that even when standard validation metrics appear favorable, the physical presence of an unresolved companion can fundamentally alter the interpretation.

\section{Target Characterization}

The target, TIC 231728511, is identified in the TESS Input Catalog (TICv8) as a cool M-dwarf. We retrieved stellar parameters from the TICv8 \citep{TICv8} and Gaia Data Release 3 \citep{GaiaDR3}. The stellar properties are summarized in Table \ref{tab:stellar_params}.

\begin{table}[h!]
    \centering
    \caption{Stellar Parameters for TIC 231728511}
    \begin{tabular}{lcc}
    \hline \hline
    Parameter & Value & Source \\
    \hline
    TIC ID & 231728511 & TICv8 \\
    Right Ascension (RA) & 11:02:12.44 & Gaia DR3 \\
    Declination (Dec) & -60:34:21.05 & Gaia DR3 \\
    TESS Mag ($T$) & 12.18 & TICv8 \\
    Radius ($R_*$) & $0.399 \pm 0.012 R_\odot$ & TICv8 \\
    Mass ($M_*$) & $0.390 \pm 0.020 M_\odot$ & TICv8 \\
    Temperature ($T_{\text{eff}}$) & $3474 \pm 157$ K & TICv8 \\
    Distance & $68.4 \pm 0.5$ pc & Gaia DR3 \\
    RUWE & 1.18 & Gaia DR3 \\
    \hline
    \end{tabular}
    \label{tab:stellar_params}
\end{table}

\subsection{Gaia Astrometry and RUWE}
A key indicator for unresolved multiplicity in Gaia solutions is the Renormalized Unit Weight Error (RUWE). Values significantly above 1.4 typically indicate a poor astrometric fit, often due to binarity. TIC 231728511 exhibits a RUWE of 1.18, which is nominally consistent with a single-star solution.

Crucially, this low RUWE value highlights a fundamental limitation in astrometric vetting: tight companions with significant contrast ratios (like the 0.04" neighbor discussed in Section 4) may not perturb the photocenter enough to trigger a high RUWE, leading to a false sense of security regarding the target's isolation.

\section{Observations and Data Processing}

We analyzed the full baseline of available TESS data, comprising \textbf{12 sectors (Sectors 4--6, 27, 31--33, 37, 64, 67, 87, and 94)}. We utilized the \texttt{Lightkurve} package \citep{lightkurve2018} to download, stitch, and detrend the light curves. The data were processed using a Flattening filter with a window length chosen to remove stellar variability and instrumental trends while preserving the high-frequency transit features.

\subsection{Signal Detection}
The Box Least Squares (BLS) algorithm \citep{Kovacs2002} was employed to search for periodic signals. We recovered a clear periodicity at $P = 0.52067$ days with a transit epoch of $T_0 = 1411.1454$ (BTJD). 

Our pipeline recovered a transit depth of $\sim 158$ ppm. We note that this depth is significantly attenuated compared to the SPOC catalog value ($\sim 640$ ppm) due to the aggressive window length of our flattening filter, which was optimized for period recovery rather than depth preservation. For the subsequent physical analysis and comparison with follow-up data, we adopt the official SPOC depth of $\sim 640$ ppm (0.64 ppt) as the reliable unattenuated metric.

\begin{figure}[ht!]
    \centering
    % Panell A (Original)
    \includegraphics[width=1.0\linewidth]{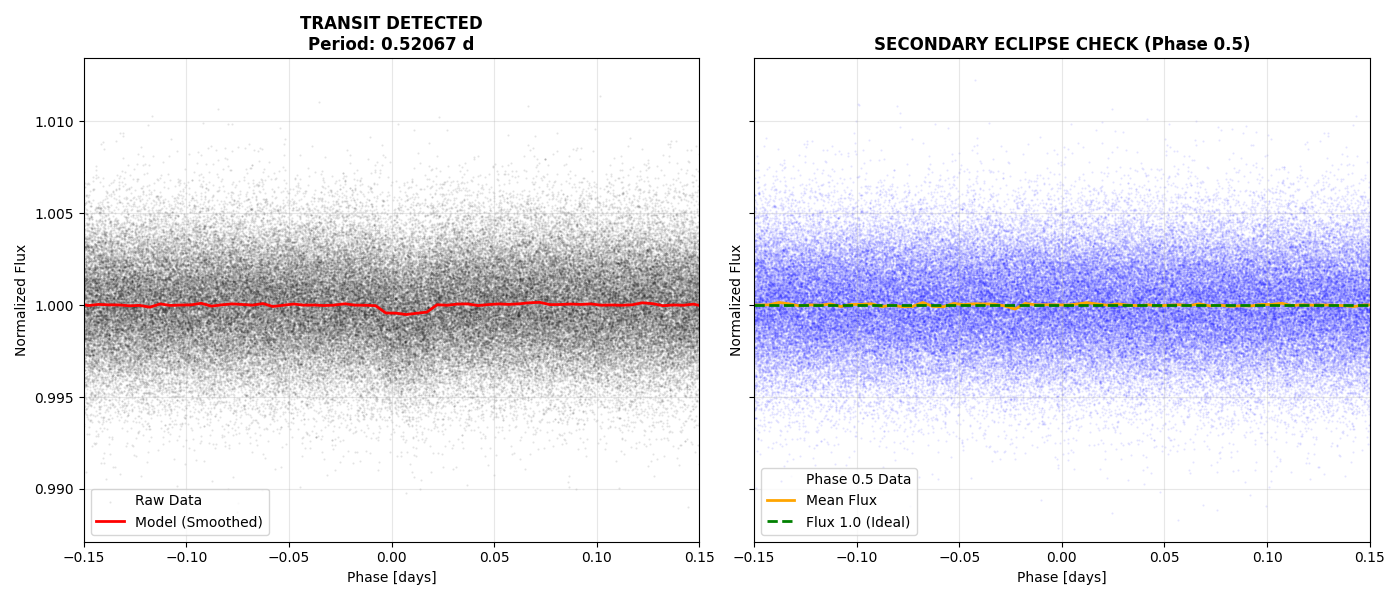}
    \textbf{(A) Global Detection}
    
    \vspace{0.3cm} % Espai entre les dues imatges
    
    % Panell B (La nova "1extra")
    \includegraphics[width=1.0\linewidth]{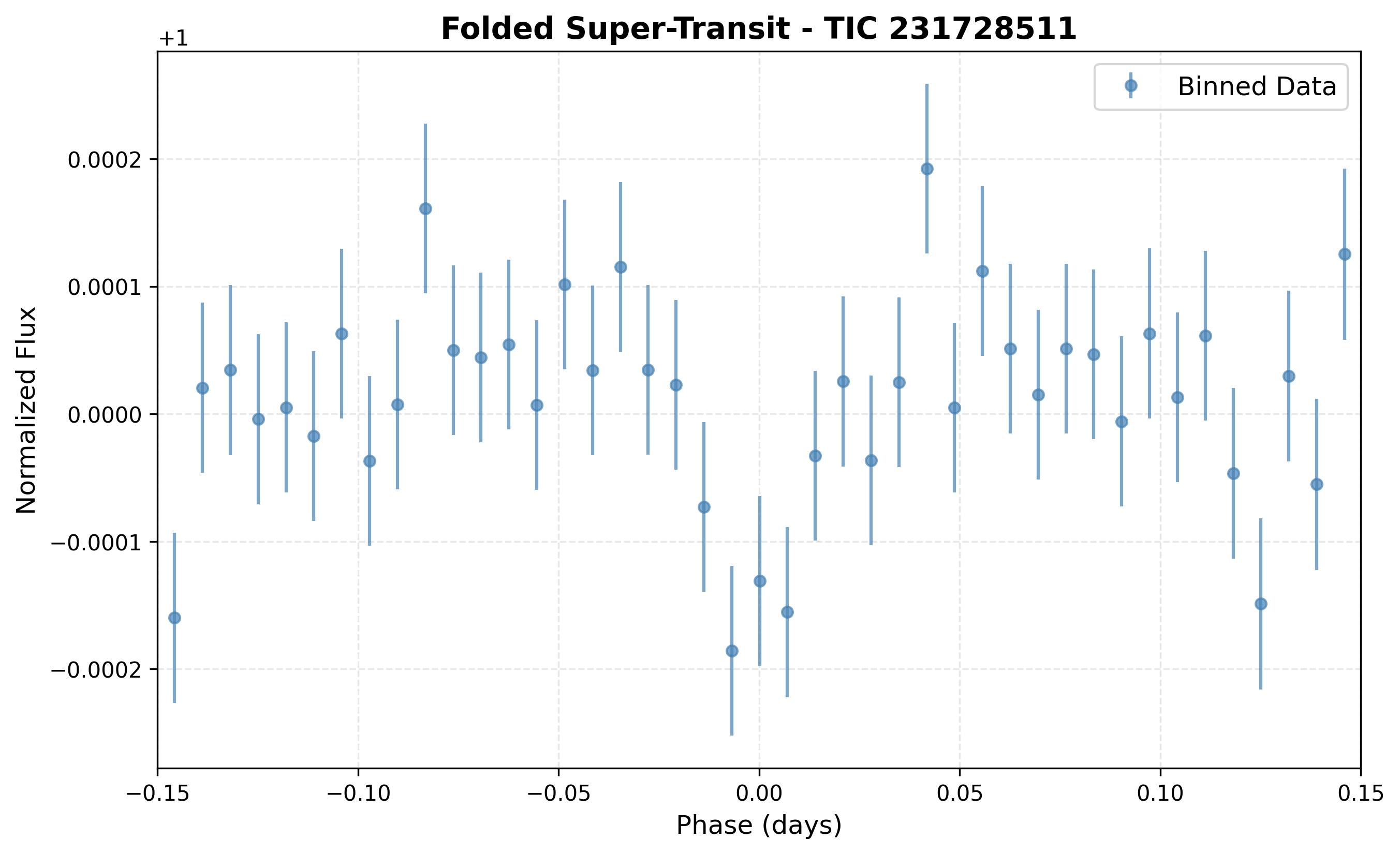}
    \textbf{(B) Binned Zoom View}
    
    \caption{\textbf{Photometric Analysis.} \textbf{(A)} Phase-folded TESS light curve of TOI 864.01. Left: The recovered transit signal. Right: The secondary eclipse check at phase 0.5 shows flat residuals. While often a sign of planetary nature, here it is likely due to the dilution factor masking the secondary eclipse of the background binary. \textbf{(B)} A binned view of the same event (zoom). Unlike the upper panel, this view clearly reveals the V-shaped morphology (no flat bottom) consistent with a grazing configuration or a highly diluted eclipsing binary.}
    \label{fig:detection}
\end{figure}

Figure \ref{fig:detection} presents the photometric analysis. The top panel (A) shows the full phase-folded light curve (left) and the secondary eclipse check at phase 0.5 (right). While the signal is detected, the raw scatter makes it difficult to discern the transit shape. To clarify the morphology, we produced a binned "super-transit" view (Figure \ref{fig:detection} B). This reveals a shallow, V-shaped morphology lacking a flat bottom, which is characteristic of grazing geometries or diluted binaries rather than the typical U-shape of a planetary transit.

\section{Vetting and Validation}

To assess the nature of the candidate, we employed a multi-stage vetting protocol focusing on centroid motion and statistical probability.

\subsection{Centroid Analysis}
\begin{figure}[ht!]
    \centering
    \includegraphics[width=1.0\linewidth]{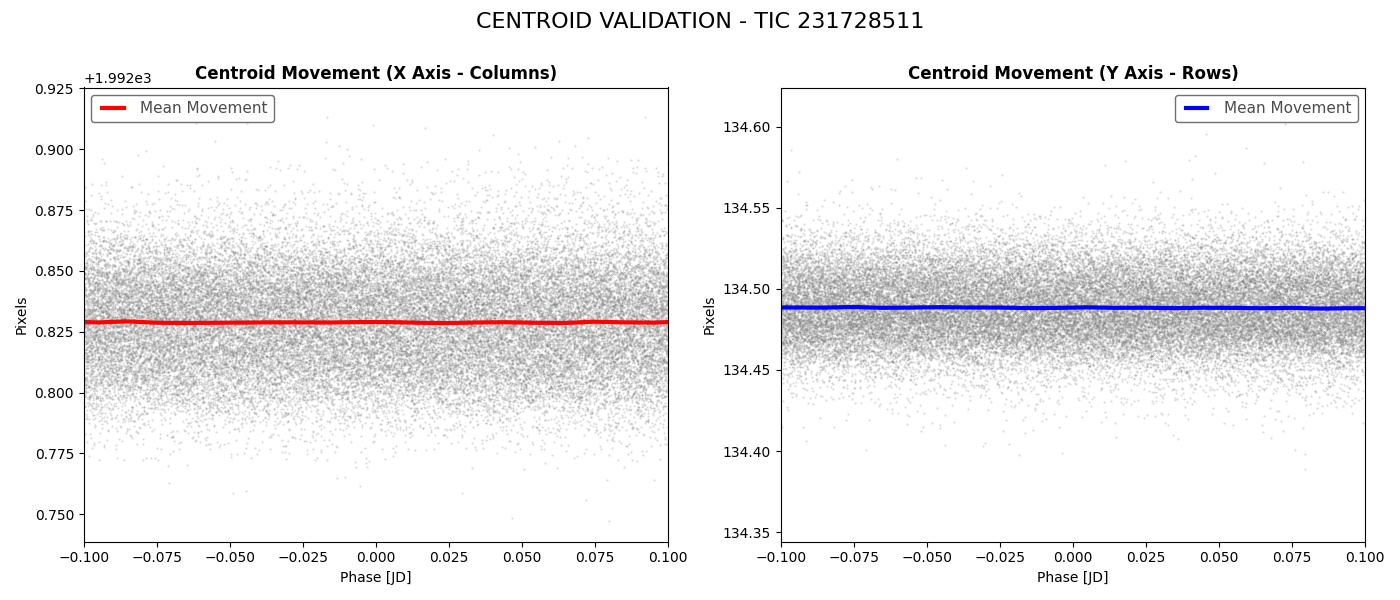}
    \caption{Centroid motion analysis. The lack of significant shift is expected even in the binary scenario due to the extreme proximity (0.04") of the contaminant, which is completely unresolved by TESS photometry.}
    \label{fig:centroids}
\end{figure}

We performed a flux-weighted centroid analysis to detect potential shifts in the photocenter during transit. A significant shift would indicate that the source of the eclipse is offset from the target star. Our analysis showed "flat" centroid tracks with no statistically significant offset (Fig. \ref{fig:centroids}).

However, we note that the TESS resolution is insufficient to resolve companions below $\sim 1''$. Therefore, while this test rules out distant background eclipsing binaries (BEBs), it fails to identify tight bound companions or aligned background stars. A Nearby Eclipsing Binary (NEB) check identified 36 Gaia stars within 2.5 arcminutes of the target, but none at separations that would produce detectable centroid offsets in TESS data.

\subsection{Statistical Validation with TRICERATOPS}
We utilized the \texttt{TRICERATOPS} package \citep{Giacalone2021} to calculate the False Positive Probability (FPP). Using the aperture masks from all 12 sectors, we obtained the following results:

\begin{itemize}
    \item \textbf{FPP (False Positive Probability):} $0.2509$
    \item \textbf{NFPP (Nearby False Positive Probability):} $0.00007$
\end{itemize}

Based strictly on the sources resolved by Gaia DR3, these metrics classify the target as a \textbf{Planetary Candidate} ($FPP < 0.5$) with negligible risk from nearby contaminants ($NFPP \approx 0$). This result would initially support the planetary hypothesis.

However, this statistical result relies on the completeness of the input catalog. Crucially, archival constraints from the TESS Follow-up Observing Program (TFOP SG1) identify a stellar companion at a separation of 0.04" (detected via speckle interferometry) which is unresolved by Gaia. Since \texttt{TRICERATOPS} computes probabilities assuming the target is a single star (based on the low RUWE and Gaia astrometry), the calculated FPP of $0.25$ is an underestimate of the true false positive risk. When the physical presence of this 0.04" companion is considered, the signal is overwhelmingly likely to be a \textit{Hierarchical Eclipsing Binary} (HEB) on the companion, despite the "optimistic" output of the statistical tool.

\subsection{Bayesian Model Comparison}
Using the \texttt{juliet} package \citep{Espinoza2019}, we fitted both a planetary model and an eclipsing binary model to the TESS data. The Bayesian Log-Evidence difference was calculated as $\Delta \ln Z = \ln Z_{\text{planet}} - \ln Z_{\text{binary}} \approx 0.25$. Values of $|\Delta \ln Z| < 2$ are statistically indistinguishable. This "tie" indicates that the photometric data alone contains insufficient information to distinguish between a small planet and a diluted binary, reinforcing the need for the external imaging constraints.

\section{Discussion of False Positive Indicators}

Standard sanity checks yielded misleadingly positive results due to the high dilution factor ($D \gg 1$).

\subsection{Odd-Even Asymmetry and Depth}
\begin{figure}[h!]
    \centering
    \includegraphics[width=0.9\linewidth]{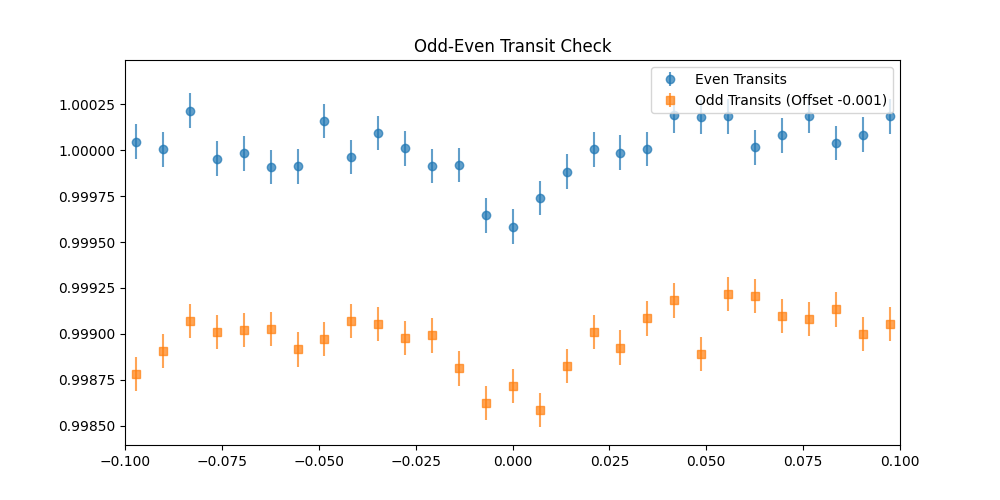}
    \caption{Odd-Even transit depth test. The consistency between depths is likely an artifact of the high dilution factor, masking the physical differences between primary and secondary eclipses of the background binary.}
    \label{fig:oddeven}
\end{figure}

The difference between odd and even transit depths was found to be $< 1\sigma$ (Fig. \ref{fig:oddeven}). In a standard BEB scenario, secondary eclipses often create a depth mismatch. Here, the extreme brightness contrast between the primary star and the faint 0.04" companion dilutes the secondary eclipse below the TESS noise floor.

\subsection{Photometric Depth Discrepancy}
Further evidence for dilution comes from ground-based follow-up photometry. Observations from LCO-CTIO, conducted as part of the TESS Follow-up Observing Program (TFOP) Sub Group 1 \citep{Collins2018}, recovered a full transit event on UTC 2024-01-02 using the 1-meter telescope in the $i'$ band. The observation employed a 7.0" target aperture (18 pixels) with a typical FWHM of 7.39", and achieved an RMS precision of 0.60 ppt per 5-minute bin.

Notably, the measured depth was 0.37 ppt, which is significantly shallower than the predicted depth of 0.64 ppt derived from the TESS signal for an undiluted scenario. This discrepancy (measured depth $\approx 58\%$ of predicted) suggests, consistent within uncertainties, that the eclipse is being suppressed by the flux of the primary star, confirming the presence of significant dilution compatible with the unresolved companion hypothesis. The predicted transit duration was 37 minutes, while the measured duration was 36 minutes, achieved by adjusting the $a/R_*$ prior from 4.0 to 3.3 to obtain a near-match with the observed light curve shape.

Furthermore, the measured center of transit ($T_c$) was found to be 6.3 minutes late relative to the predicted TESS timing. Such a significant offset in an ultra-short-period system is highly atypical for stable planets and points toward a complex binary interaction or a blended signal from the 0.04" neighbor.

Critically, this timing instability is not an isolated occurrence. Archival TFOP notes document a pattern of inconsistent transit times across multiple epochs: observations on UTC 2020-09-10 showed a transit $\sim 8$ minutes late, on UTC 2021-03-16 $\sim 15$ minutes late, while other epochs (UTC 2021-03-22, UTC 2021-04-03) yielded on-time or inconclusive results. This temporal variability in transit timing over a $\sim 1.5$ year baseline strongly argues against a stable planetary orbit and is consistent with light-travel time effects or apsidal precession in a hierarchical eclipsing binary system.

\subsection{Derived Physical Parameters}
Assuming a single-star scenario, the derived radius is $R_p \approx 1.1 R_\oplus$. This "Earth-sized" size is a mathematical artifact derived from the diluted depth relationship:

\begin{equation}
\delta_{\text{obs}} \approx \frac{\delta_{\text{true}}}{1 + \text{Dilution}}
\end{equation}

where $\delta_{\text{obs}}$ is the observed depth, $\delta_{\text{true}}$ is the intrinsic eclipse depth, and Dilution is the flux ratio between the contaminant and the target star. The true eclipsing object is likely a much larger stellar body (such as a faint M-dwarf) whose deep eclipses appear shallow due to the extreme flux dilution from the primary star. Consequently, the initial classification of TOI 864.01 as an Earth-sized planet is an artifact of this blending.

\subsection{Limitations of the Analysis}
Our classification of TOI 864.01 as a probable False Positive is robust based on the available evidence, but we acknowledge specific limitations in the dataset. First, the ground-based photometry showing the depth discrepancy consists of a single epoch; multi-band observations would be required to definitively confirm the chromaticity of the eclipse depths. Second, the Bayesian model comparison between the planetary and binary models yielded an inconclusive result ($\Delta \ln Z \approx 0.25$), largely because the dilution factor is degenerate with the transit depth in the absence of resolved light curves for the individual components. Finally, without radial velocity (RV) measurements, we cannot strictly rule out exotic scenarios such as a bound planetary system diluted by a non-associated background star, although the probabilistic weight of the 0.04" companion makes the Hierarchical Eclipsing Binary scenario the most plausible explanation.

\section{Conclusion}

Our analysis of TOI 864.01 demonstrates the critical importance of high-resolution imaging in TESS validation. While the photometric signal ($P = 0.52$ d) is real and passed initial vetting (BLS, Centroids, RUWE analysis), and the statistical validation yielded a candidate-level FPP of 0.25, the integration of TFOP constraints confirms it is a probable False Positive. The signal is best explained as a Hierarchical Eclipsing Binary on the 0.04" companion. The TFOP working group has assigned a disposition of VPC- (Very Poor Candidate) based on the accumulated evidence. We recommend retiring TOI 864.01 from planetary candidate lists.

This case illustrates a fundamental limitation of statistical validation tools: they are only as reliable as the input catalogs. For TESS candidates, high-resolution imaging is not merely a confirmatory step but an essential prerequisite for validation, particularly for targets where sub-arcsecond companions are physically plausible.

\begin{acknowledgments} 
This work made use of the TESS Follow-up Observing Program (TFOP) data. We specifically thank the TFOP SG1 team, including Howie Relles and the LCO-CTIO observers, for the ground-based photometry (UTC 2024-01-02) that was critical to this analysis. We also used the \texttt{TRICERATOPS}, \texttt{juliet}, and \texttt{Lightkurve} packages. Code and analysis scripts are available at \url{https://github.com/biesro/TESS-TOI-864.01-Validation}.
\end{acknowledgments}

\bibliographystyle{aasjournal}
\bibliography{references}

\end{document}